\documentclass[12pt]{thesis}

\usepackage[dvips]{graphicx} 
\usepackage{amsfonts}
\usepackage{amscd}
\usepackage{amsmath}    

\title{\textbf{Security of Quantum Key Distribution with Realistic Devices}}
\author{{\large \textbf{Xiongfeng Ma}}\\
\\
A report submitted in conformity with the requirements for Master of Science\\
\\
Department of Physics\\
\\
University of Toronto\\
\\
Supervisor: Hoi-Kwong Lo\\}

\begin{document}

\maketitle

\begin{abstract}
We simulate quantum key distribution (QKD) experimental setups and
give out some improvement for QKD procedures. A new data
post-processing protocol is introduced, mainly including error
correction and privacy amplification. This protocol combines the
ideas of GLLP and the decoy states, which essentially only requires
to turn up and down the source power. We propose a practical way to
perform the decoy state method, which mainly follows the idea of
Lo's decoy state. A new data post-processing protocol is then
developed for the QKD scheme with the decoy state. We first study
the optimal expected photon number $\mu$ of the source for the
improved QKD scheme. We get the new optimal $\mu=O(1)$ comparing
with former $\mu=O(\eta)$, where $\eta$ is the overall transmission
efficiency. With this protocol, we can then improve the key
generation rate from quadratic of transmission efficiency
$O(\eta^2)$ to $O(\eta)$. Based on the recent experimental setup, we
obtain the maximum secure transmission distance of over 140 km.
\end{abstract}

\chapter{Introduction} Quantum key distribution (QKD) \cite{BB84,
Ekert} allows two parties, commonly called Alice, the transmitter
and Bob, the receiver, to create a random secret key with the
channel revealed to the eavesdropper, Eve. The security of QKD is
built on the fundamental laws of physics in contrast to existing key
distribution schemes that are based on unproven computational
assumptions.

The best-known QKD scheme was proposed by Bennett and Brassard in
1984 (commonly called BB84 protocol) \cite{BB84}. Alice uses a
quantum channel, which is governed by quantum mechanics to transit
single photons, each in one of four polarizations: horizontal
($0^{\circ}$), vertical ($90^{\circ}$), $45^{\circ}$, $135^{\circ}$.
Bob randomly chooses one of two bases: rectangular ($+$) or diagonal
($\times$) to measure the arrived signals, and keeps the result
privately. Consequently, Alice and Bob compare the bases they use
and discard those in different bases. At last, Alice and Bob perform
the local operations and classical communications (LOCC) to do the
data post-processing, which is mainly composed of error correction
and privacy amplification \cite{DEJM,LoChau,ShorPr}. Our report will
improve the procedure of BB84 with decoy state \cite{HwangDecoy},
\cite{LoDecoy} and apply the idea of GLLP \cite{GLLP} to develop a
new post-processing protocol.

The security of the idealized QKD system has been proven in the past
few years \cite{Mayers,LoChau,ShorPr}. Now let us turn our attention
to the experiment. Real setup is no longer ideal, but with imperfect
sources, noisy channels and inefficient detectors, which will affect
the security of the QKD system.

A weak coherent state is commonly used as the photon source, which
is essentially a mixture of states with Poisson distribution of
photon number. Thus, there is a non-zero probability to get a state
with more than one photon. Then Eve may suppress the quantum state
and keep one photon of the state that has more than one photon
(commonly called multi photon, correspondingly, single photon
denotes the state with only one photon). Moreover, Eve may block the
single photon state, split the multi photon state and improve the
transmission efficiency with her superior technologies to compensate
the loss of blocking single photon. Without the technology to
identify the photon numbers, Alice and Bob have to pessimistically
assume all the states lost in the transmission and detection are
single photons. In this way, the secure expected photon number $\mu$
of the signal state is roughly given by, $\mu=O(\eta)$, which
implies that the key generation rate $R=O(\eta^2)$, details can be
found in section \ref{PriorArt} and Appendix A.

The inefficiency of the detector will also affect the security of
the QKD system. There exists a so-called dark count in the realistic
detectors, which will increase the error rate of detection
especially when the transmission efficiency is low. The \emph{dark
count} of a detector denotes the probability to get detection events
when there is no input to the detector. Eve may use this
imperfection to cover the error she introduces from her measurement
of the single photon state.

In the recent paper GYS \cite{GYS}, the authors conducted an
experiment in which the single photon was transmitted over 120 km.
The question is whether the QKD experiment reported in GYS is secure
or not? Unfortunately, based on the prior art of post-processing
scheme, it is insecure. GYS uses the expected photon number
$\mu=0.1$ as the source, which will be defeated by so-called photon
number splitting attack in long distance. All in all, there exists a
gap between the theory and experiment. Here, we are attempting to
bridge the theory and experiment.

The key problem here is that Bob does not know whether his detection
events come from: single photon, multi photon, or dark count. To
solve this problem, we apply the idea of decoy state to learn the
performance of single photon states. The decoy state here acts as a
``scope" telling Alice and Bob which state comes from single photon,
multi photon or dark count.

Our result is significant because it is a bridge between the theory
and experiment of QKD. we extend the secure distance of the QKD
system, increase the key generation rate substantially and maintain
the major advantage of QKD --- unconditional security. We improve
the key generation rate from $O(\eta^2)$ to $O(\eta)$. Notice that
such a key generation rate is the highest order that any QKD system
can achieve. Our improvement is mainly based on an advanced theory.
We do not require any enhancement of equipment but only turning up
and down the source power, which is easy to implement with current
technology.

We notice that Koashi \cite{koashi} has also proposed a method to
extend the distance of secure QKD. It will be interesting to compare
the power and limitations of our approach with that of Koashi.

The outline of this report is as follows. In section \ref{EDP} we
will recall to the proof of the security of an idealized QKD system
with entanglement distillation protocol (EDP). In section
\ref{Simulation}, we shall review a couple of widely used QKD setups
and simulates them following \cite{Lutkenhaus}. In section
\ref{PriorArt}, we shall investigate the former post-processing
schemes with the simulation and point out the limitation of prior
art. We find out that the key generation rate $R$ is $O(\eta^2)$,
where $\eta$ is the overall transmission efficiency. In section
\ref{Decoy}, we combine the idea of GLLP \cite{GLLP} and decoy state
\cite{HwangDecoy, LoDecoy}, and improve the key generation rate from
$O(\eta^2)$ to $O(\eta)$. In Appendix A, we discuss the choosing of
the optimal expected photon number $\mu$, which maximizes the key
generation rate. In Appendix B, we introduce a practical way to
perform weak decoy state, which is a key step of the decoy state
method.


\chapter{EDP schemes for QKD} \label{EDP}
In this section we will recall the security proof of the idealized
QKD with EDP \cite{BBPS, DEJM, BDSW, LoChau}. The security of BB84
scheme can be reduced to the security of the Entanglement
Distillation Protocol (EDP) schemes \cite{ShorPr}.

In the EDP protocol, Alice creates $n+m$ pairs of qubits, each in
the state
$$|\psi\rangle=\frac{1}{\sqrt2}(|00\rangle+|11\rangle),$$
the eigenstate with eigenvalue $1$ of the two commuting operators
$X\bigotimes X$ and $Z\bigotimes Z$, where
$$X=\begin{pmatrix}
    0 & 1\\
    1 & 0\\
    \end{pmatrix},
  Z=\begin{pmatrix}
    1 & 0\\
    0 & -1\\
    \end{pmatrix}$$
are the Pauli operators. Then she sends half of each pair to Bob.
Alice and Bob sacrifice $m$ randomly selected pairs to test the
error rates in the $X$ and $Z$ bases by measuring $X\bigotimes X$
and $Z\bigotimes Z$. If the error rate is too high, they abort the
protocol. Otherwise, they conduct the EDP, extracting $k$
high-fidelity pairs from the $n$ noisy pairs. Finally, Alice and Bob
both measure $Z$ on each of these pairs, producing a \textit{k-bit}
shared random key about which Eve has negligible information. The
protocol is secure because the EDP removes Eve's entanglement with
the pairs, leaving her negligible knowledge about the outcome of the
measurements by Alice and Bob.

A QKD protocol based on a CSS-like EDP can be reduced to a
``prepare-and-measure" protocol \cite{ShorPr}. CSS code \cite{CSS}
conducts error correction to the bit error and the phase error
separately. That is to say, CSS-like EDP deals with the error
correction and privacy amplification separately, which can be
further improved by two-way communication \cite{GL}. Thus the
residue of this data post-processing protocol is the so-called CSS
rate,
\begin{equation}\label{EDP:CSSrate}
\eta_{post}^{CSS}=1-H_2(\delta_b)-H_2(\delta_p)
\end{equation}
where $\delta_b$ and $\delta_p$ are the bit flip error rate and the
phase flip error rate, and $H_2(x)$ is binary Shannon information
function,
$$
H_2(x)=-x\log_2(x)-(1-x)\log_2(1-x).
$$

In summary, there are two main parts of EDP, bit flip error
correction (for error correction) and phase flip error correction
(for privacy amplification). These two steps can be understood as
follows. First Alice and Bob apply bi-direction error correction,
after which they share the same key strings but Eve may still keep
some information about the key. Alice and Bob then perform the
privacy amplification to expunge Eve's information from the key. The
final key will be secure if the privacy amplification is
successfully done in principle. That is Alice and Bob do not need to
actually do the privacy amplification but only ensure that this can
be done. In practice, Alice and Bob can calculate the residue of the
privacy amplification and perform random hashing to get the final
key with high security.


\chapter{Simulations for experiments}\label{Simulation}
To simulate a real-life QKD system, we need to model the source,
channel and detector. In this section, we first review the real
experimental setup, and then simulate the QKD system following
\cite{Lutkenhaus}, at last verify the simulation with real
experimental data.

\section{QKD setup}
\begin{figure}[hbt]
\centering \resizebox{8cm}{!}{\includegraphics{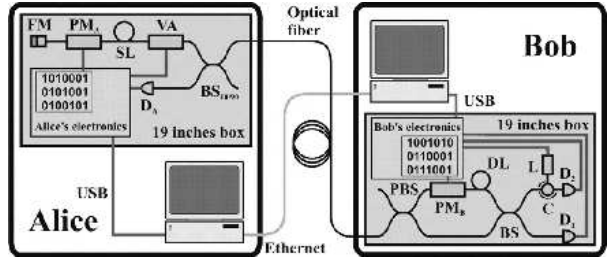}}
\caption{Schematic of the p\&p prototype. This figure comes from
\cite{NJP67}} \label{Simulation:fig:ppsetup}
\end{figure}

Let us recall the principle of the so-called p\&p auto-compensating
setup \cite{MHHT, NJP67}, where the key is encoded in the phase
between two pulses trading from Bob to Alice and back (see Fig.
\ref{Simulation:fig:ppsetup}). A strong laser pulse emitted from Bob
is separated by a first 50/50 beam splitter (BS). The two pulses
impinge on the input ports of a polarization beam splitter (PBS),
after having traveled through a short arm and a long arm, including
a phase modulator ($PM_B$) and a delay line (DL), respectively. All
fibers and optical elements at Bob are polarization maintaining. The
linear polarization is rotated by $90^\circ$ in the short arm,
therefore the two pulses exit Bob's setup by the same port of the
PBS. The pulses travel down to Alice, are reflected on a Faraday
mirror, attenuated, and come back orthogonally polarized. In turn,
both pulses now take the other path at Bob and arrive at the same
time at the BS where they interfere. Then, they are detected either
in $D_1$, or after passing through the circulator (C) in $D_2$.
Since the two pulses take the same path, inside Bob in reversed
order, this interferometer is auto-compensated.\cite{NJP67}

To implement the BB84 protocol, Alice applies a phase shift of $0$
or $\pi$ and  $\pi/2$ or $3\pi/2$ on the second pulse with $PM_A$.
Bob chooses the measurement basis by applying a $0$ or $\pi/2$ shift
on the first pulse on its way back.\cite{NJP67}

As for the free-space QKD system, the setup is easier. There are
only encoded signals which come from Alice's side to Bob's,
comparing with the p\&p setup, the pulses traveling from Bob to
Alice and back. In this sense, the free-space setup is closer to the
original BB84 schemes. More details of this kind of QKD setup can be
found in \cite{BHKL,BHLM,KZHW}.

\section{Modeling the real-life QKD system}\label{Modeling}
Following prior papers such as \cite{Lutkenhaus}, we simulate the
p\&p QKD system. The first important parameter for the real-life QKD
system is the key bit rate $B$ between Alice and Bob. More
explicitly, $B$ is the number of exchanged key bits \emph{per
second}, given by
\begin{equation}\label{Simulation:Bit}
B=\nu R,
\end{equation}
where $\nu$ is the repetition frequency in Alice's side, and $R$ is
the key generation rate, i.e. the number of exchange bits \emph{per
pulse}, given by
\begin{equation}\label{Simulation:Rate1}
R=q p_D \eta_{post},
\end{equation}
where $q$ depends on the implementation (1/2 for the BB84 protocol,
because half the time Alice and Bob disagree with the bases, and if
one uses the efficient BB84 protocol \cite{EfficientBB84}, one can
have $q\approx1$), $p_D$ is the average number of signals per pulse
detected by Bob, and $\eta_{post}$ denotes for the residue of data
post-processing, i.e. the efficiency of error correction and privacy
amplification, which is also discussed in section \ref{EDP} Eq.
\eqref{EDP:CSSrate}. We will study $p_D$ in the following and
discuss $\eta_{post}$ in the section \ref{PriorArt} and \ref{Decoy}.

For optical fibers, the losses in the quantum channel can be derived
from the loss coefficient $\alpha$ measured in dB/km and the length
of the fiber $l$ in km. The channel transmission $t_{AB}$ can be
expressed as
$$
t_{AB}=10^{-\frac{\alpha l}{10}}.
$$
Let $\eta_{Bob}$ denote for the internal transmission $t_{Bob}$ and
detection efficiency $\eta_D$ in Bob's side, given by
$$
\eta_{Bob}=t_{Bob}\eta_D.
$$
Then the overall transmission and detection efficiency between Alice
and Bob $\eta$ is given by
\begin{equation}\label{Simulation:Eta}
\eta=t_{AB}\eta_{Bob}.
\end{equation}

The average number of signals per pulse detected by Bob, i.e. the
probability for Bob to get a signal from his detector $p_D$, is
given by
\begin{equation}\label{Simulation:pD1}
\begin{aligned}
p_D&=p_{Signal}+p_{dark}-p_{Signal}p_{dark}\\
   &\cong p_{Signal}+p_{dark},
\end{aligned}
\end{equation}
where we assume that the dark counts are independent of the signal
photon detection. $p_{dark}$ and $p_{Signal}$ are the probabilities
to get a dark count and to detect a photon originally emitted by
Alice respectively.

The dark count depends on the characteristics of the photon
detectors. The effect of dark count will be significant when
$\eta\mu$ is small which implies $p_{Signal}$ is small. It is
necessary to point out that $p_{dark}$ is the overall dark count
throughout the QKD system. Here we consider the p\&p QKD setup,
$p_{dark}$ is twice as large as $d_B$ because there are two sources
of dark count, i.e. two detectors in the QKD system. Then the
$p_{dark}$ is given by
$$
p_{dark}=2d_B,
$$
where $d_B$ denotes the dark count of one detector.

Normally, a weak coherent state is used as the signal source.
Assuming that the phase of this signal is totally randomized, the
number of photons of the signal state follows a Poisson distribution
with a parameter $\mu$ as its expected photon number. In appendix B,
we will discuss the optimal value of expected photon number $\mu$
which optimizes the key generation rate $R$. $p_{Signal}$ is given
by
\begin{equation}\label{Simulation:pSignal}
p_{Signal}=\sum_{i=1}^{\infty}\eta_i\cdot\frac{\mu^i}{i!}\exp(-\mu),
\end{equation}
where $\eta_i$ is the transmission efficiency of \emph{i-photon}
state in a normal channel. It is reasonable to assume independence
between the behaviors of the $i$ photons. Therefore the transmission
efficiency of \emph{i-photon} state $\eta_i$ is given by
\begin{equation}\label{Simulation:etai}
\eta_i=1-(1-\eta)^i.
\end{equation}
Substitute \eqref{Simulation:etai} into \eqref{Simulation:pSignal},
we have,
\begin{equation}\label{Simulation:pSignal1}
\begin{aligned}
p_{Signal}&=\sum_{i=1}^{\infty}[1-(1-\eta)^i]\cdot\frac{\mu^i}{i!}\exp(-\mu)\\
&=1-\exp(-\eta\mu).
\end{aligned}
\end{equation}

We can divide $p_{Signal}$ into two parts $p_S$ and $p_M$, which are
the probabilities of single photon and multi photon states emitted
from Alice's side that are detected by Bob. Then
\eqref{Simulation:pD1} is given by,
\begin{equation}\label{Simulation:pD}
\begin{aligned}
p_D&=p_{dark}+p_S+p_M\\
&=p_{dark}+1-e^{-\eta\mu}.\\
\end{aligned}
\end{equation}

The overall quantum bit error rate (QBER, denoted by $\delta$) is an
important parameter for error correction and privacy amplification
$\eta_{post}$. QBER is equivalent to the ratio of the probability of
getting a false detection to the total probability of detection per
pulse. It comes from three parts: dark count, single photon and
multi photon,
\begin{equation}\label{Simulation:QBER1}
\begin{aligned}
\delta&=\frac{\frac12p_{dark}+\delta_S p_S+\delta_M p_M}{p_D},
\end{aligned}
\end{equation}
where $\delta_S$ and $\delta_M$ denote the error rate of single and
multi photon detection, respectively. The dark counts occur
randomly, thus the error rate of dark count is $\frac12$.

Due to the high loss in the channel, the multi photon states
arriving at Bob's side always has only one photon left. Thus, we
have the similar probability of erroneous detection for single
photon and multi photon. The bit error rate $\delta_S$ and
$\delta_M$ are given by,
\begin{equation}\label{Simulation:deltaSM}
\delta_M\cong\delta_S=e_{detector},
\end{equation}
where $e_{detector}$ is the probability that a photon hit the
erroneous detector. $e_{detector}$ characterizes the quality of the
optical alignment of the polarization maintaining components and the
stability of the fiber link \cite{NJP67}. It can be measured with
strong pulses, by always applying the same phases and measuring the
ratio of the count rates at the two detectors.  In our discussion,
we neglect $e_{detector}$'s dependence of the fiber length because
the change of alignment during the transmission of a signal is very
small.

Substituting \eqref{Simulation:deltaSM} into
\eqref{Simulation:QBER1}, QBER is given by,
\begin{equation}\label{Simulation:QBER}
\delta=\frac{\frac12 p_{dark} + e_{detector}p_{Signal}}{p_D}.
\end{equation}

There is another important parameter for privacy amplification, the
ratio of single photon detection in overall detection events $f_1$
which is defined by,
\begin{equation}\label{Simulation:f1}
f_1=\frac{p_S}{p_D}.
\end{equation}
Only the key extracted from the single photon state can be secure.
Thus, $f_1\cdot p_D$ roughly gives the upper bound of the key
generation rate.

\section{Verify the simulation by QBER}
Here we would like to verify the equations \eqref{Simulation:QBER}
by comparing the experimental result and our simulation. The
parameters of the experimental setup are listed in Tab.
\ref{Simulation:Table:data}.
\begin{table}[h]\center 
\begin{tabular}{|c|c|c|c|c|}
\hline
& T8\cite{T8} & G13\cite{G13} & KTH\cite{KTH} & GYS\cite{GYS}\\
\hline
Wavelength [nm] & 830 & 1300 & 1550 & 1550 \\
\hline
$\alpha$ [dB/km] & 2.5 & 0.32 & 0.2 & 0.21 \\
\hline
$t_B$ [dB] & 8 & 3.2 & 1 & 5* \\
\hline
$e_{detector}$ [\%]& 1 & 0.14 & 1 & 3.3 \\
\hline
$d_B$ [per slot] & $5\times10^{-8}$ & $8.2\times10^{-5}$ & $2\times10^{-4}$ & $8.5\times10^{-7}$ \\
\hline
$\eta_D$ [\%]& 50 & 17 & 18 & 12* \\
\hline
\end{tabular}
\caption{Key parameters for p\&p QKD experiment setup.
* GYS gives out that $\eta_{Bob}=0.045$.} \label{Simulation:Table:data}
\end{table}

\begin{figure}[hbt]
\centering \resizebox{8cm}{!}{\includegraphics{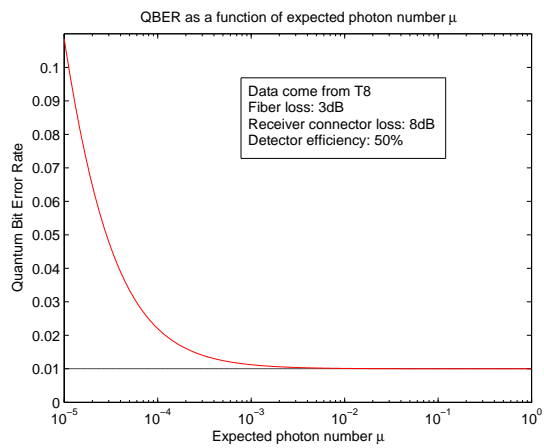}}
\caption{shows QBER as a function of input expected photon number
using Eq. \eqref{Simulation:QBER}, reproducing T8's Fig 3. The key
parameters are, according to T8 \cite{T8}, listed in Tab.
\ref{Simulation:Table:data}, and the fiber loss is 3dB
($\sim1.5$km).} \label{Simulation:fig:VerifyT8}
\end{figure}

Fig. \ref{Simulation:fig:VerifyT8} shows QBER as a function of
expected photon number $\mu$. For $\mu$ valuing in the range
$10^{-1}\sim10^{-4}$, the QBER has a constant value of about $1\%$
that is dominated by depolarization-induced errors $e_{detector}$.
These errors arise from dynamic depolarization that results from the
finite extinction ratios of the various polarizing components in the
system \cite{T8}. For $\mu<10^{-4}$ the QBER rises as the dark
counts falling within the detectors start to become the dominant
contribution to the noise.

\begin{figure}[hbt]
\centering \resizebox{8cm}{!}{\includegraphics{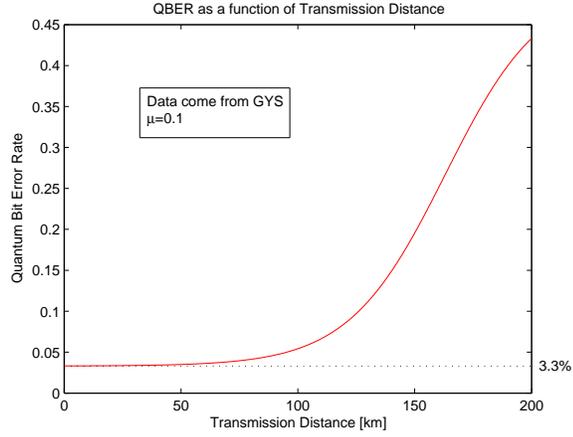}}
\caption{shows QBER as a function of the transmission distance,
using the Eq. \eqref{Simulation:QBER}. This is a reproduction of
FIG.3 in GYS' paper \cite{GYS}. The key parameters are, according to
GYS \cite{GYS}, listed in Tab. \ref{Simulation:Table:data}. The
expected photon number $\mu$ is 0.1.}
\label{Simulation:fig:VerifyGYS}
\end{figure}

Fig. \ref{Simulation:fig:VerifyGYS} shows QBER as a function of
transmission distance using Eq. \eqref{Simulation:QBER}. In the long
distance ($l>100km$, say), the QBER rises as the dark counts become
the dominant contribution to the noise. Note the exponential
dependence is due to the loss of photons in the propagation,
referring to $p_{Signal}$ of Eq. \eqref{Simulation:pSignal1}. If
stronger source (say, $\mu=0.5$) is used, QBER will be lower,
especially in the long distance region.

From these verifications, we can see that the Eq.
\eqref{Simulation:QBER} fits the experiment well, so the simulation
is accurate.


\chapter{Our comparison of Prior Art Results} \label{PriorArt}
Now, we begin to examine the relationship between key generation
rate $R$ and the transmission distance with prior QKD data
post-processing schemes. Two ideas in L\"{u}tkenhaus' paper
\cite{Lutkenhaus} and GLLP's paper \cite{GLLP} are compared. Note
that it is our new work to calculate the key generation rate $R$
with the experimental simulation using the idea of GLLP \cite{GLLP}.

In both papers, the authors are going on the assumption that all of
the photons that fail to arrive were emitted as single photons.
Later, we will improve this point by introducing the decoy states
scheme. Assuming that,
\begin{equation} \label{PriorArt:pmsm}
p_M=S_M,
\end{equation}
where $S_M$ is the probability of emitting a multi photon state from
Alice's side, which is dependent on attribute of the source. Based
on the Poisson distribution of the number of photons of the signal
states,
$$
S_M=1-(1+\mu)\exp(-\mu).
$$
Another assumption used here is that all the error (QBER) comes from
the single photon state. Then the error rate of single photon is
given by,
\begin{equation} \label{PriorArt:deltaS}
\delta_S=\frac{\delta}{f_1}.
\end{equation}

For individual attacks, the residue of error correction and
privacy amplification can be given by \cite{Lutkenhaus},
\begin{equation}\label{PriorArt:LutRate}
\begin{aligned}
\eta_{post}=&\max\{f_1(1-\log_2[1+\frac{4\delta}{f_1}-4(\frac{\delta}{f_1})^2])\\
&-f(\delta)H_2(\delta),0\},
\end{aligned}
\end{equation}
where $f(\delta)\ge1$ is the efficiency of error correction
\cite{BS} listed in Tab. \ref{PriorArt:Table:fe}, $\delta$ is the
QBER given in equation \eqref{Simulation:QBER}, $f_1$ is defined in
\eqref{Simulation:f1}.

\begin{table}[hbt]
\center 
\begin{tabular}{|c|c|c|c|c|} \hline
$\delta$ & 0.01 & 0.05 & 0.1 & 0.15 \\
\hline
f($\delta$) & 1.16 & 1.16 & 1.22 & 1.35\\
\hline
\end{tabular}
\caption{\normalfont{The data come from \cite{Lutkenhaus}. The
author used used the upper bounds for I(4) provided in \cite{BS}.}}
\label{PriorArt:Table:fe}
\end{table}

We would like to explain the idea of GLLP's \cite{GLLP} tagged state
briefly here. Notice the idea of tagged state is (perhaps
implicitly) introduced by \cite{ILM}. In principle, one can separate
the tagged and untagged states, i.e. one can do random hashing for
the privacy amplification on the tagged state and untagged state
separately. Therefore, the data post-processing can be performed as
following. First, apply error correction to the overall states,
sacrificing a fraction $H_2(\delta_b)$ of the key, which is
represented in the first term of formula \eqref{PriorArt:GLLPRate}.
After correcting errors in the sifted key string, one can imagine
executing privacy amplification on two different strings, the sifted
key bits $s_{tagged}$ arising from the tagged qubits and the sifted
key bits $s_{untagged}$ arising from the untagged qubits. Since the
privacy amplification \cite{GLLP} is linear (the private key can be
computed by applying the $C_2$ parity check matrix to the sifted key
after error correction), the key obtained is the bitwise $XOR$
$$
s_{untagged}\oplus s_{tagged}
$$
of keys that could be obtained from the tagged and untagged bits
separately. If $s_{untagged}$ is private and random, then it doesn't
matter if Eve knows everything about $s_{tagged}$ --- the sum is
still private and random. Therefore we ask if privacy amplification
is successful applied to the untagged bits alone. Thus, the residue
after error correction and privacy amplification can be expressed
as,
\begin{equation}\label{PriorArt:GLLPRate}
\eta_{post}=\max\{-f(\delta)H_2(\delta)-f_1(1-H_2(\frac{\delta}{f_1})),0\}.
\end{equation}
Here the single photon state is regarded as the untagged state and
its error rate ($\delta_b=\delta_p$) is given by
\eqref{PriorArt:deltaS}.

Based on \eqref{Simulation:Rate1}, \eqref{PriorArt:LutRate} and
\eqref{PriorArt:GLLPRate}, according to Appendix B, the optimal
expected photon number $\mu$, which maximizes the key generation
rate, is roughly given by,
\begin{equation} \label{PriorArt:OptMu}
\mu \approx\eta,
\end{equation}
where $\eta$ is the overall transmission, defined in the
\eqref{Simulation:Eta}. Therefore, the key generation rate is given
by,
\begin{equation} \label{PriorArt:quadratic}
R=O(\eta\mu)=O(\eta^2).
\end{equation}

Now, use \eqref{PriorArt:OptMu} as the expected photon number to
calculate the key generation rate of two different schemes by
equation \eqref{PriorArt:LutRate} and \eqref{PriorArt:GLLPRate} with
L\"{u}tkenhaus' and GLLP's post-processing protocols.

\begin{figure}[hbt]
\centering \resizebox{8cm}{!}{\includegraphics{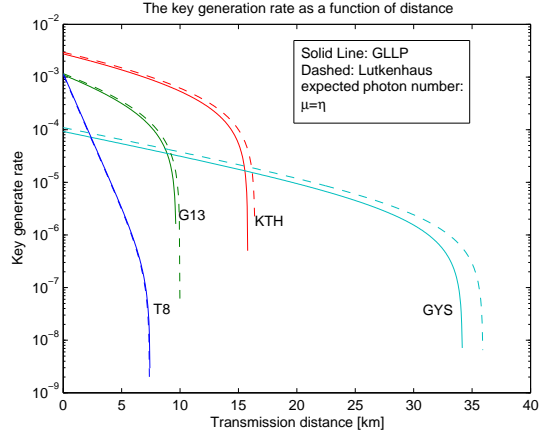}}
\caption{shows the relationship between key generation rate and the
transmission distance, comparing L\"{u}tkenhaus' individual attack
and GLLP's general attack case. The key parameters are listed in
Tab. \ref{Simulation:Table:data}. It is our new work to calculate
the key generation rate $R$ with the experiment simulation using the
idea of GLLP \cite{GLLP}.} \label{PriorArt:fig:LutGLLP}
\end{figure}

Fig. \ref{PriorArt:fig:LutGLLP} shows the relationship between key
generation rate and the transmission distance by one-way LOCC,
comparing L\"{u}tkenhaus' individual attack and GLLP's general
attack case. From the Fig. \ref{PriorArt:fig:LutGLLP}, we can see
that GLLP is only slightly worse than L\"{u}tkenhaus', but GLLP deal
with the general attack while L\"{u}tkenhaus' result is restricted
in individual attack. Our result shows that there seems to be little
to be gained in restricting security analysis to individual attacks,
given that the two papers---L\"{u}tkenhaus vs GLLP---gave very
similar results. In other words, our view is that one is better off
in considering unconditional security, rather than restricting one's
attention to a restricted class of attacks (such as individual
attacks). Note that the key generation rate $R$ of GYS with GLLP
will strictly hit 0 at distance $l=34km$.

\chapter{Decoy state method}\label{Decoy}
So far as discussed, the prior art gives Eve many ideal advantages
such as she can make the tagged state no error and no loss in the
channel, however, this is not necessary. We can control Eve's
performance on tagged states by adding decoy states. ``Control" does
not mean limit the quantum or classical computation ability of Eve,
but means we can detect Eve if she uses some eavesdropping
strategies to enforce the tagged states.

\section{The idea of decoy state}
The decoy state method is proposed by Hwang \cite{HwangDecoy} and
further studied by \cite{LoDecoy}. The idea is that, by adding some
decoy states, one can estimate the behavior of vacua, single photon
states, and multi photon states individually.

The key point is that, with the decoy state, Alice and Bob can gain
photon number information which cannot be derived by today's
technologies directly. They can use this extra information to
``detect" the behavior of states with different photon numbers. Eve
cannot distinguish whether the photon comes from signal state or
decoy state. Thus, the transmission efficiency $\{\eta_i\}$,
detection probabilities and error rates ($\delta_S$, $\delta_M$) in
the signal state will be the same as those in decoy state,
$$
\begin{aligned}
\eta_i(Signal)&=\eta_i(Decoy)\\
\delta_S(Signal)&=\delta_S(Decoy)\\
\delta_M(Signal)&=\delta_M(Decoy),
\end{aligned}
$$
where $i=1,2,3\cdots$.

Our decoy state method is quite different from Hwang's original one.
Hwang uses strong pulse as decoy state. We mainly follow
\cite{LoDecoy}'s decoy idea, using vacua and very weak state as
decoy states.

\section{Simulation with new assumption} \label{Decoy:Simulation}
With the decoy states, we can drop the ``pessimistic" assumption
\eqref{PriorArt:pmsm}. According to Poisson distribution of photon
source, the single photon and multi photon detection probabilities
are given by
\begin{equation}\label{Decoy:pS}
p_S=\eta\mu\exp(-\mu)
\end{equation}
\begin{equation}\label{Decoy:pM}
\begin{aligned}
p_M&=\sum_{i=2}^{\infty}[1-(1-\eta)^i]\frac{\mu^i}{i!}e^{-\mu}\\
&=1-e^{-\eta\mu}-\eta\mu e^{-\mu},
\end{aligned}
\end{equation}
where $\mu$ is the expected photon number and $\eta$ is the overall
transmission and detection efficiency, defined in
\eqref{Simulation:Eta}. The transmission efficiency of an {\em
i-photon} state $\eta_i$ is given by \eqref{Simulation:etai}.

We remark that the Eqs. \eqref{Decoy:pS} and \eqref{Decoy:pM} do not
consider dark count contributions. In real-life, dark count
contributions must be included. Therefore, following Eq.
\eqref{Simulation:pD1}, we write down the total contribution (true
signals plus dark counts) as follows,
\begin{equation}\label{Decoy:pBob}
\begin{aligned}
\tilde{p}_{dark}&=p_{dark}e^{-\mu} \\
\tilde{p}_S&=p_S+p_{dark} \mu e^{-\mu} \\
\tilde{p}_M&=p_M+p_{dark}[1-(1+\mu)e^{-\mu}] \\
\end{aligned}
\end{equation}
and the error rate of dark count is still $\frac12$, for single
photon and multi photon,
\begin{equation}\label{Decoy:deltaBob}
\begin{aligned}
\tilde{\delta}_S&=\frac{\frac12p_{dark}+e_{detector}p_S}{p_{dark}+p_S}\\
\tilde{\delta}_M&=\frac{\frac12p_{dark}+e_{detector}p_M}{p_{dark}+p_M}.
\end{aligned}
\end{equation}

\section{A way to perform decoy state} \label{Decoy:AWay}
Here we would like to introduce a specific method to perform decoy
state, which is proposed by \cite{LoDecoy}. There are three kind of
signals Alice and Bob should perform.

First, Alice and Bob can study the dark counts by using vacua as
decoy states. Here we reasonably assume that the density matrix of
dark counts is the \emph{identity matrix}. Thus, in theory, they
will detect the probability to get a signal $p_D^{vacua}$ and the
overall error rate $\delta^{vacua}$,
\begin{equation}\label{Decoy:Vacua}
\begin{aligned}
p_D^{vacua}&=p_{dark}=2d_B\\
\delta^{vacua}&=\frac12.
\end{aligned}
\end{equation}
They can measure $p_D^{vacua}$ and $\delta^{vacua}$ via experimental
methods and then compare with the Eq.\eqref{Decoy:Vacua}. If the
experimental results match the Eq.\eqref{Decoy:Vacua}, then keep
going. Otherwise, they need to check out the QKD system setup,
especially the detectors.

Secondly, Alice and Bob can get the transmission and the bit flip
error rate of the single photon state by using very weak coherent
states as decoy states. With weak decoy state, according to
\eqref{Simulation:pD} and \eqref{Simulation:QBER}, $p_D^{weak}$ and
$\delta^{weak}$ are given by,
\begin{equation}\label{Decoy:Weak}
\begin{aligned}
p_D^{weak}&=p_{dark}+p_S^{weak}+p_M^{weak}\\
\delta^{weak}&=\frac{\frac12p_{dark}+\delta_S p_S^{weak}+\delta_M
p_M^{weak}}{p_D^{weak}},
\end{aligned}
\end{equation}
where the superscript \emph{weak} denotes the value comes from weak
decoy state. In appendix B, we will discuss the how to choose
$\mu^{weak}$ in practice.

If the decoy state is weak enough, i.e. $\mu^{weak}\ll1$, at which
value
\begin{equation} \label{Decoy:WeakRelation}
\frac{p_M^{weak}}{p_S^{weak}}=O(\mu^{weak})\ll1,
\end{equation}
we can neglect the multi photon terms in Eq.\eqref{Decoy:Weak}.
After that we have
$$
p_S^{weak}=p_D^{weak}-p_{dark},
$$
and substitute this formula into \eqref{Decoy:pBob},
\eqref{Decoy:deltaBob},
\begin{equation}
\begin{aligned}
\tilde{p}_S^{weak}&=p_D^{weak}-p_{dark}(1-\mu^{weak}e^{-\mu^{weak}}) \\
&\cong p_D^{weak}-p_{dark}e^{-\mu^{weak}}\\
\tilde{\delta}_S&\cong\frac{\delta_{weak}p_D^{weak}-\frac12p_{dark}e^{-\mu^{weak}}}{p_S^{weak}}
\end{aligned}
\end{equation}
where $p_{dark}$ can be derived from the Eq.\eqref{Decoy:Vacua}, and
$p_D^{weak}$, $\delta_{weak}$ can be obtained from the experiment.

Thirdly, Alice and Bob perform the signal state, where the final key
is drawn from. The detection probability $p_D$ and QBER $\delta$ are
given by \eqref{Simulation:pD} and \eqref{Simulation:QBER}.

\section{Data post-processing}
Here, we would like to discuss the data post-processing for QKD with
the decoy state, mainly based on \cite{GLLP}.

We can further extend the GLLP's idea \cite{GLLP}, as discussed in
section \ref{PriorArt}, to more than one kind of tagged states case,
i.e. several kinds of states with flag $g$. The procedure of data
post-processing is similar, do the overall error correction first
and then apply the privacy amplification to each case. At last the
residue of error correction and privacy amplification is given by,
\begin{equation}\label{Decoy:RateGeneral}
\eta_{post}=\max\{-f(\delta_b)H_2(\delta_b)+\sum_g
p^g[1-H_2(\delta_p^g)],0\}
\end{equation}
where one need to sum over all cases with flag $g$, $p^g$ is the
probability of the case with flag $g$ and $\delta_p^g$ is the phase
flip error rate of the state with flag $g$

In order to combine the ideas of GLLP and Decoy, we should find out
all parameters for Eq. \eqref{Decoy:RateGeneral}.  There are three
kind of states in the discussion, dark count, single photon, and
multi photon. The multi photon state emitting out of the source can
be expressed by $|000\rangle+|111\rangle$. After Eve gets the extra
photon, the state will be a mixture of $|00\rangle+|11\rangle$ and
$|00\rangle-|11\rangle$ with the same probability. Thus the phase
flip error rate for multi photon will be $\frac12$. In addition to
the discussion in \ref{Decoy:Simulation}, we can list the bit flip
error rate $\delta_b$ and phase flip error rate $\delta_p$ for dark
count, single photon and multi photon in Tab.
\ref{Decoy:Table:delta}.
\begin{table}[hbt]
\center 
\begin{tabular}{|c|c|c|c|} \hline
& Dark count & Single Photon & Multi Photon \\
\hline
$\delta_b$ & $\frac12$ & $\tilde{\delta}_S$ & $\tilde{\delta}_M$ \\
\hline
$\delta_p$ & $\frac12$ & $\tilde{\delta}_S$ & $\frac12$ \\
\hline
\end{tabular}
\caption{Bit flip error rate and phase flip error rate for different
kinds of photon state.} \label{Decoy:Table:delta}
\end{table}

Apply above idea, Eq. \eqref{Decoy:RateGeneral}, to decoy states
scheme with the parameters listed in Tab. \ref{Decoy:Table:delta}.
We get the formula for post-processing residue,
\begin{equation}\label{Decoy:Postpro}
\eta_{post}=\max\{-f(\delta_b)H_2(\delta_b)+\frac{\tilde{p}_S}{p_D}[1-H_2(\tilde{\delta}_S)],0\}
\end{equation}
where $\tilde{p}_S$ and $\tilde{\delta}_S$ are given by
\eqref{Decoy:pBob}, \eqref{Decoy:deltaBob}. Note that the dark count
and multi photon state have no contribution to the final key with
phase flip error rate $\delta_p=\frac12$. The reason why we use
$\tilde{p}_S$ and $\tilde{\delta}_S$ here is that Bob cannot
distinguish the detection event from real signal or dark count. And
substitute \eqref{Decoy:Postpro} into the equation
\eqref{Simulation:Rate1}, if the key generation rate is $R>0$,
\begin{equation}\label{Decoy:KeyRate}
R=\frac12\{-p_Df(\delta)H_2(\delta)+\tilde{p}_S[1-H_2(\tilde{\delta}_S)]\}
\end{equation}
where $\delta$ is the overall QBER given in \eqref{Simulation:QBER}.

Through the analysis in Appendix B, we have, for the KTH \cite{KTH}
experimental setup,$\mu_{Optimal}\approx0.8$ and for GYS \cite{GYS},
$\mu_{Optimal}\approx0.5$. Therefore, the key generation rate is
given by,
\begin{equation} \label{Decoy:linear}
R=O(\eta\mu)=O(\eta)
\end{equation}

\begin{figure}[hbt]
\centering \resizebox{8cm}{!}{\includegraphics{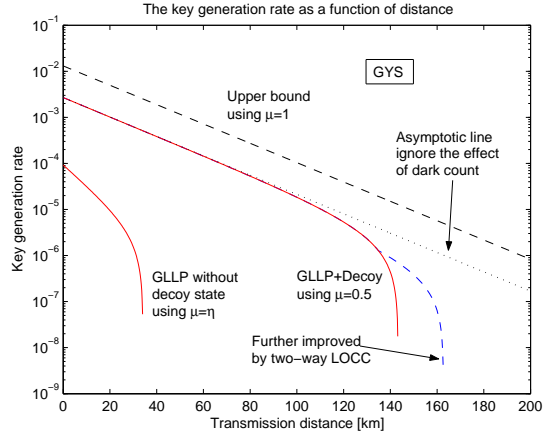}}
\caption{shows the key generation rate as a function of the
transmission distance, GLLP+Decoy Eq. \eqref{Decoy:KeyRate}. The key
parameters are listed in Tab. \ref{Simulation:Table:data} and f(e)
is given by Tab. \ref{PriorArt:Table:fe}. We use linear regression
for f(e).} \label{Decoy:fig:Decoy+GLLP}
\end{figure}

The result is shown in Fig.\ref{Decoy:fig:Decoy+GLLP}.

\noindent Remarks for the figure:
\begin{enumerate}
\item The dashed line is the upper bound of the key generation
rate. The key generation rate here is derived from the photon
detection events of Bob that occur when Alice sends single photon
signals. It is to say, Bob can distinguish dark count, single
photon, and multi photon. In this way the upper bound of key
generation rate is given by,
$$
R_{max}=Q_1(1-H_2(e_{detector})).
$$
It is obvious that the optimal expected photon number for the upper
bound is $\mu=1$. The gap between the upper bound and the decoy
curve shows how much room is left for improvements of data
post-processing.

\item The dotted line is the asymptotic line that neglects the
influence of dark count, which is given by,
$$
\begin{aligned}
R_{Asym}=&-p_Df(e_{detector})H_2(e_{detector})\\
&+p_S[1-H_2(e_{detector})]\\
p_D=&1-\exp(\eta\mu),
\end{aligned}
$$

\item The ``decoy" and ``without decoy" line will hit zero at some
points. This can be seen from the formula \eqref{Decoy:KeyRate},
$R=0$ when,
$$
\noindent
\begin{aligned}
&-Q_{Signal}f(E_{Signal})H_2(E_{Signal})\\
&+Q_1[1-H_2(\delta_S)] \le 0.
\end{aligned}
$$
From the program (as shown in figure), when $l=144km$ for
GLLP+Decoy, and $l=34km$ for GLLP, $R$ will hit 0.

\item One can further improve the data post-processing protocol with
two-way LOCC \cite{GL}. As shown in Fig.\ref{Decoy:fig:Decoy+GLLP},
dashed curve outside of decoy curve, two-way LOCC improve the
maximum distance by about 20 km.
\end{enumerate}

Compare the curves with and without decoy, we can find that the
advantages of decoy state are obvious,
\begin{enumerate}
\item  The initial key generation rate (at zero distance) is
substantially higher with decoy state than without. It is because a
stronger source (higher $\mu$) is used in the decoy state scheme.

\item   At short distances, the key generation rate decreases with
distance exponentially mainly due to exponential losses in the
channel. The two curves, with and without decoy state behave like
straight lines. Note that the initial slope without decoy is twice
of that with decoy state. This is because without decoy state,
$R=O(\eta^2)$, while with decoy state, $R=O(\eta)$ according to
\eqref{PriorArt:quadratic} and \eqref{Decoy:linear}.

\item  Suppose $10^{-6}$ is the cut-off point for key generation rate.
The two curves intercept the cut-off point at rather different
locations (with decoy: 139 km, and without decoy: 31 km). For GYS,
the distance is over 100km. It is comparable to the distance between
amplifiers in optical metropolitan area networks (MANs).
\end{enumerate}

Here, we would like to discuss the key bit rate $B$, which is
different from behavior with key generation rate $R$ only with a
constant, repetition frequency, according to Eq.
\eqref{Simulation:Bit}. It is necessary to point out that there are
two repetition frequencies in real-life setups: the source frequency
$\nu_A$, which is the limitation of signal source repetition at
Alice's side and the detection rate $\nu_B$, which is the limitation
of detector's count rate in Bob's side. There are two cases we
should consider. First, when the overall transmission loss is not
too large, the detection frequency $\nu_B$ limits the final key bit
rate, because we can complement the loss in the transmission through
increasing the source frequency $\nu_A$ until reaches $\nu_A$'s
maxima. In this case, $B$ is simply given by, $B=\nu_B$. Secondly,
when the overall transmission loss is large, $B$ is determined by
the source rate $\nu_A$. Then the key bit rate $B$ is given by
\eqref{Simulation:Bit}, $B=\nu_A R$.

We remark that the decoy state idea can also be used for free-space
QKD setup \cite{Free23}. The simulation of the setup will be
similar.


\chapter*{Conclusion}
In this report, we have presented a security
proof of quantum cryptography against the general attack with
real-life devices. We have formulated a method for estimating the
key generation rate in the presence of realistic devices by
combining the idea of decoy state and GLLP. The model has been
applied to various real-life experimental setups. Based on recent
experiment results, secure transmission distance of the QKD system
can be up to 140 km.

We have improved the key generation rate substantially. The main
reason for this improvement is that with decoy state we can choose
the expected photon number $\mu$ of source in the order of O(1),
while without decoy state, $\mu=O(\eta)$. With the decoy state, a
new data post-processing protocol is developed, which essentially
comes from the idea of GLLP. We have also proposed a pratical way to
fulfills the ideal of decoy state as discussed in section
\ref{Decoy:AWay} and Appendix B. In this sense, it can help to
design the QKD procedure.

Through the simulation, it is clearly shown that, in order to
improve the transmission distance, one should reduce the dark count
and fiber loss in the channel; as for the aim of higher key bit
rate, we should increase the detection repetition or reduce the
errors in the transmission.


\chapter*{Acknowledgments}
First I would like to thank Professor Hoi-Kwong Lo's patient
supervision. His direction is highly insightful and supportive. In
the beginning, I modeled the experimental setup following the
simulation of \cite{Lutkenhaus}. The discussion with Norbert
L\"{u}tkenhaus is really helpful. Then I compared GLLP and
L\"{u}tkenhaus's results with this simulation. I was dissatisfied
with the poor maximum distance for the QKD system. I was struggling
to ``control" Eve's behavior until Hoi-Kwong gave me his paper
\cite{LoDecoy} about the decoy states. After I got this paper, I
found that decoy state was the right way to ``control" Eve's
behavior. More strictly, decoy state is a good way to ``detect"
behavior of state with different photon numbers through the channel.
All the figures are produced by the programs written in MatLab. I
highly appreciate that Kai Chen reproduced all my programs and
checked many versions of my report. I discussed with Bing Qi about
the experimental data and setup and asked the authors of experiment
\cite{GYS} directly. I discussed about two-way LOCC with Daniel
Gottesman. All of the above researchers are very supportive to my
work. I thank helpful discussions with various colleagues including,
Kai Chen, Daniel Gottesman, Norbert L\"{u}tkenhaus, and Bing Qi.
Finally, I would like to thank Ryan Bolen for his proof reading of
this report.

This report has been merged with \cite{LoDecoy} into a new paper
\cite{LMC}.


\chapter{Appendix A}
In this section, we will discuss the choosing of the expected photon
number $\mu$ for different error correction and privacy
amplification schemes in section \ref{PriorArt}, \ref{Decoy}. We
will discuss the optimal $\mu$ generally and then work out
reasonable value for each scheme.

We would like to start with generic discussion. On one hand, we need
to maximize the probability of single photon detection, which is the
only source for the final key. To achieve this point, we should
maximize the single photon sources since transmission efficiency is
fixed. Considering the real photon sources, according to Poisson
distribution of the photon number, the single photon source reaches
its maximum when $\mu=1$. On the other hand, we have to control the
probability of multi photon detection to ensure the security of the
system. On this side, we should keep the tagged states ratio
($1-f_1$) small, which requires $\mu$ not too large. It follows that
$\mu\le1$, because based on both points, the case of $\mu>1$ is
always worse than the case of $\mu=1$, And another parameter should
be considered is the QBER $\delta$, which is decrease when $\mu$
increases, according to the formulas \eqref{Simulation:QBER}.
Therefore, intuitively we have that,
$$
\mu \in (0,1].
$$

\section{Without decoy state}
Here, we would like to consider the case without decoy state, i.e.
GLLP and L\"{u}tkenhaus's cases. A similar discussion is given in
\cite{Lutkenhaus}. We desire to get an optimal value of $\mu$ that
maximizes the key generation rate $R$ with other parameters fixed.
The key parameters here are the overall transmission and detection
efficiency $\eta$, dark count $d_B$, and the probability of
erroneous detection $e_{detector}$, which are specified by various
setups.

In the \emph{R-distance} figures, such as Fig.
\ref{PriorArt:fig:LutGLLP} and \ref{Decoy:fig:Decoy+GLLP}, the key
generation rate drops roughly exponentially with the transmission
distance before it starts to drop faster due to the increasing
influence of the dark counts. The initial behavior is mainly due to
the multi-photon component of the signals while the influence of the
error-correction part is small. In this regime we can bound the gain
by the approximation
$$
\begin{aligned}
R & \le \frac12(p_{Signal}-p_M)\\
& = \frac12[(1+\mu)\exp(-\mu)-\exp(-\eta\mu)]
\end{aligned}
$$
with the pessimistic assumption \eqref{PriorArt:pmsm}. This
expression is optimized if we choose $\mu=\mu_{Optimal}$, which
fulfills
$$
-\mu\exp(-\mu)+\eta\exp(-\eta\mu)=0.
$$
Since for a realistic setup we expect that $\eta\mu \ll 1$, we find
\begin{equation} \label{AppendixB:PriorMu}
\eta_{Optimal}\approx\eta.
\end{equation}

\begin{figure}[hbt]
\centering
\resizebox{8cm}{!}{\includegraphics{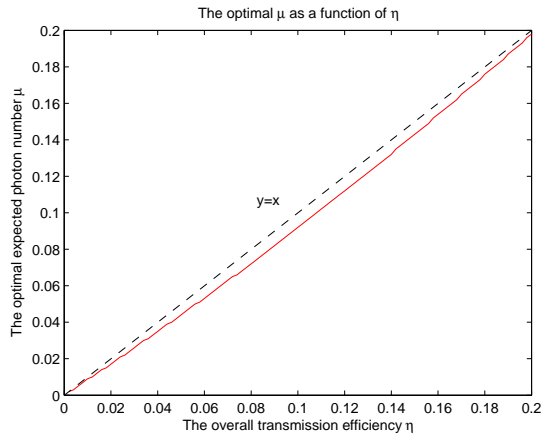}}\caption{Optimal
expected photon number $\mu$ as a function of transmission
efficiency $\eta$, with the parameters listed in Tab.
\ref{Simulation:Table:data}, T8 \cite{T8}. Here, we use dichotomy
method to search the region to get the optimal $\mu$ that maximizes
the key generation rate \eqref{Simulation:Rate1} and
\eqref{PriorArt:GLLPRate}.} \label{AppendixB:fig:Prior}
\end{figure}

Now, we can use the numerical analysis to verify the formula
\eqref{AppendixB:PriorMu}. When we keep all parameters fixed and
vary the expected photon number $\mu$ of the signal, then we can use
dichotomy method to find out the $\mu_{Optimal}$, which maximizes
the key generation rate by the formula \eqref{Simulation:Rate1} and
\eqref{PriorArt:GLLPRate}. If we fix the dark count $d_B$ and the
probability of erroneous detection $e_{detector}$, and vary the
transmission efficiency $\eta$ we can draw the relationship between
the optimal $\mu_{Optimal}$ and $\eta$. The result is shown in Fig.
\ref{AppendixB:fig:Prior}, from which we can clearly see that the
formula \eqref{AppendixB:PriorMu} is a good approximation.

\section{With decoy state}
In principle, with the decoy state, we can control the performance
of tagged states. So $\mu_{Optimal}$ should maximize the untagged
states ratio $f_1$, as defined in \eqref{Simulation:f1}. Thus,
$\mu_{Optimal}$ should be greater than \eqref{AppendixB:PriorMu}.

If we keep all parameters fixed and vary the expected photon number
of the signal, we can obtain a key generation rate curve with a
clear maximum. The key generation rate is given by
\eqref{Decoy:KeyRate}.

\begin{figure}[hbt]
\centering \resizebox{8cm}{!}{\includegraphics{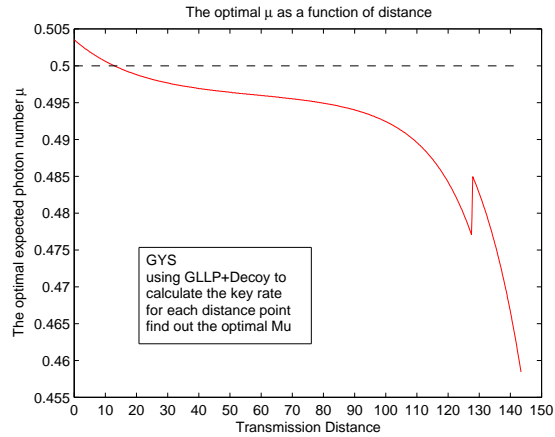}}
\caption{Using numerical analysis to obtain the optimal key
generation rate by the formula \eqref{Decoy:KeyRate} and use the
data GYS in Tab. \ref{Simulation:Table:data}, with parameters listed
in Tab. \ref{Simulation:Table:data}.} \label{AppendixB:fig:DecoyGen}
\end{figure}

We would like to start with numerical analysis on
\eqref{Decoy:KeyRate} directly. For each distance we find out the
optimal $\mu$ that maximizes the key generation rate. The result is
shown in Fig. \ref{AppendixB:fig:DecoyGen}. The strange behavior of
the curve around $l=125km$ is due to the linear regression of error
correction efficiency f(e) given in Tab. \ref{PriorArt:Table:fe}.
Without f(e), the curve will be smooth. We can see that the optimal
$\mu$ for GYS is around 0.5.

Now, we would like to do analytical discussion under some
approximation. We neglect the dark count, regard $\eta\ll1$, and
consider the ideal error correction efficiency $f(e)=1$. Then
\eqref{Decoy:pBob}, \eqref{Decoy:deltaBob}, \eqref{Simulation:pD},
\eqref{Simulation:QBER} will be reduced to,
$$
\begin{aligned}
\tilde{p}_S&\cong p_S=\eta\mu e^{-\mu}\\
\tilde{\delta}_S&\cong \delta_S=e_{detector}\\
p_D&\cong p_{Signal}=1-e^{-\eta\mu}\\
\delta&\cong e_{detector}\\
\end{aligned}
$$
Substitute these formulas into Eq. \eqref{Decoy:KeyRate}, the key
generation rate is given by,
$$
R\approx\frac12\{-\eta\mu H_2(e_{detector})+\eta\mu
e^{-\mu}[1-H_2(e_{detector})]\}
$$
The expression is optimized if we choose $\mu=\mu_{Optimal}$ which
fulfills,
$$
(1-\mu)\exp(-\mu)=\frac{H_2(e_{detector})}{1-H_2(e_{detector})}.
$$
Then we can solve this equation and obtain that,
$$
\begin{aligned}
\mu_{Optimal}^{KTH} &\approx 0.8\\
\mu_{Optimal}^{GYS} &\approx 0.5
\end{aligned}
$$
where for KTH setup $e_{detector}=1\%$, and for GYS
$e_{detector}=3.3\%$. Comparing two results we can see that
numerical and analytical analysis are compatible.


\chapter{Appendix B}
In section \ref{Decoy:AWay} we use a weak decoy state. Here we would
like to discuss how to perform decoy state method in practice.

In decoy method, we require the weak decoy state to be ``weak
enough" i.e. its expected photon number $\mu\ll1$. We can then
neglect the multi photon term by considering the relationship
\eqref{Decoy:WeakRelation}. However, it is not practical for real
experiment since it will take a long time to get enough information
from weak decoy state if $\mu\ll1$.

Here we propose one possible solution for weak decoy state, using
several (say $m$) weak decoy states with expected photon number
$\mu_1,\mu_2, \cdot\cdot\cdot \mu_m$, instead of one. As discussed
in \eqref{Decoy:Vacua}, one can estimate the dark count and its
error rate accurately. Let us turn our attention to $m$ weak decoy
states. With the same argument of \eqref{Decoy:Weak}, we have,
\begin{equation} \label{AppendB:pDj}
\begin{aligned}
p_D^{\mu_j}&=p_{dark}+\sum_{i=1}^{\infty}\eta_i\cdot\frac{\mu_j^i}{i!}\cdot e^{-\mu_j}\\
\end{aligned}
\end{equation}
where $j=1,2,\cdot\cdot\cdot,m$ denotes for the \emph{j-th} weak
decoy state. To solve Eq. \eqref{AppendB:pDj}, we can neglect the
high order $(>m)$ terms which are of $O(\mu^{m+1})$, and then Eq.
\eqref{AppendB:pDj} are reduced to,
\begin{eqnarray}
\begin{aligned}
p_D^{\mu_1}&=\sum_{i=0}^{m}\eta_i\cdot\frac{\mu_1^i}{i!}\cdot e^{-\mu_1}\\
p_D^{\mu_2}&=\sum_{i=0}^{m}\eta_i\cdot\frac{\mu_2^i}{i!}\cdot e^{-\mu_2}\\
\cdot\cdot\cdot\\
p_D^{\mu_m}&=\sum_{i=0}^{m}\eta_i\cdot\frac{\mu_m^i}{i!}\cdot e^{-\mu_m}\\
\end{aligned}
\end{eqnarray}
Now, one can solve $m$ equations for $\{\eta_i\}$,
$i=1,2,3,\cdot\cdot\cdot,m$. The subsequent procedure is the same as
the \ref{Decoy:AWay}. One can use $\eta=\eta_1$ to calculate
$\tilde{p}_S$ and $\tilde{\delta}_S$ by \eqref{Decoy:pBob} and
\eqref{Decoy:deltaBob}. At last substitute the parameters into
\eqref{Decoy:KeyRate} to calculate the key generation rate.

How many decoy states should be applied, i.e. $m=?$, depends on how
low expected photon number $\mu_{weak}$ one can tolerate. Here, we
would like to give out a couple of examples. Given that
$\eta=10^{-4}$, suppose that all $m$ decoy states are in the same
order. a) If one chooses $\mu_{weak}=O(10^{-3})$ and $m=2$, then the
terms neglected are of $O(10^{-9})$ and $p_S$ is in the term of
$O(\eta\mu)=O(10^{-7})$. Then it is reasonable to neglect the multi
photon terms. b) If one chooses $\mu_{weak}=O(10^{-2})$ and $m=3$,
then the high order terms neglected are of $O(10^{-8})$ and $p_S$ is
in the term of $O(\eta\mu)=O(10^{-6})$. Then, one obtains $p_S$ and
$\delta_S$ with precision of around $1\%$.



\end{document}